**Optical humidity sensor based on a liquid whispering-gallery mode resonator**.

*Lucía Labrador-Páez[1], Kevin Soler-Carracedo[1], Miguel Hernández-Rodríguez[1], Inocencio R. Martín[1,2], Tal Carmon[3] and Leopoldo L. Martin[3,*]*.

[1]Departamento de Física, Universidad de La Laguna, San Cristóbal de La Laguna, Santa Cruz de Tenerife, 38206, Spain.
[2]MALTA Consolider Team and Instituto de Materiales y Nanotecnología (IMN), Universidad de La Laguna, San Cristóbal de La Laguna, Santa Cruz de Tenerife, 38200, Spain.
[3]Technion - Israel Institute of Technology, Haifa, 32000, Israel.
[*] Corresponding.

Email: llmartin@tx.technion.ac.il

We experimentally demonstrate the use of a novel liquid state, whispering-gallery-mode optical resonator as a highly sensitive humidity sensor. The optical resonator used consists of a droplet made of glycerol, a transparent liquid that enables high optical quality factor, doped with rhodamine 6G, which acts as fluorescent emitter. As glycerol is highly hygroscopic, the refractive index and radius of the droplet change with ambient humidity. This produces a shift on the whispering gallery mode's wavelengths, which modulates the emission of rhodamine 6G. This easily-made device has an unpreceded sensitivity of $10^{-3}$ per relative humidity percent.

While solid resonators were widely investigated, the pioneering research in WGM started with liquid droplets, [1] [2] but they were typically pumped by pulsed lasers. Recent progress in continuous wave pumped droplet resonators [3] [4] [5] allowed observation of narrow and stable optical resonances. Therefore, we can benefit from the inherent difference of liquids to improve the sensing capabilities. Ambient humidity sensing, unlike temperature or pressure, requires not only a physical change in the sensor itself, but the diffusion of water molecules inside the sensor material and since the diffusion coefficient of water in miscible liquids is orders of magnitude larger than in silica, liquid droplets are excellent candidates to develop relative humidity (RH) sensors. RH is the most commonly used parameter to quantify ambient humidity, and is defined as the percentage of water vapor present in air over the amount of water vapor needed to saturate an equal volume of air at a given temperature.[6] Furthermore, the control of RH is extremely important in many industrial processes and scientific experiments. For that reason, many devices have been proposed in order to satisfy these requirements.[6, 7] Ideally, a humidity sensor should show high sensitivity, operate along a large range of RH, react quickly, and be tough and easily-made.

Among optical sensors, those based on whispering gallery mode (WGM) resonances have the highest sensitivity. These sensors consist of optical microcavities made out of a transparent material of refractive index higher than their surroundings, hence light is confined by total internal reflection with minimal losses, due to the natural smooth surface of the liquids; therefore the surface scattering losses are negligible and high optical quality factors (Q) are achieved.[3] [8] [9] [10] [4] [11] When the refractive index and/or the radius of the optical resonator changes, the resonant wavelengths are shifted. Taking advantage of this fact, WGM optical sensors are commonly used as ambient conditions sensors [12] [13].

Most WGM sensors consist of solid resonators.[14] To the best of our knowledge, there is only one preceding work concerning liquid resonators used as optical WGM sensors,[10] where the WGM sensor was used to verify the chemical purity of diverse oils. In liquid resonators, there is full interaction between the substance to be sensed (water, in our case) and the circulating light. As a result, the sensitivity of the WGM optical sensors is higher than on solid resonators, in which only the evanescent tail of the optical modes (about 1 % of the energy) interacts with the outer medium. Moreover, this type of sensors show excellent sensitivity at all RH ranges including the extremes of RH (0-10 % and 90-100 %), where most sensors perform poorly [7, 15].

In most cases, optical humidity sensors give as output parameter the variation of power registered at the exit of an optical fiber or waveguide optically coupled to the resonator.[6, 16]

For instance, Zhang et al. have demonstrated the use of non-resonating hydrogel spheres coupled to an optical fiber core as RH optical sensor, due to the fact that the refractive index of hydrogel changes with the ambient humidity, the light is scattered out from the core of the fiber in a different amount and, consequently, the tapered fiber transmittance changes.[16] In a similar way, the spectral displacement of WGMs can be observed as a function of humidity in nanoparticle coated solid resonators, where a layer of water is adsorbed by analyzing the transmittance of a tapered fiber evanescently coupled to the resonator, in which, only 0.5% of the circulating light interacts with the water in the outer layer.[17] This technique, in addition, has two drawbacks: it requires a tunable narrow-line laser and a tapper fiber whose transmission decays when exposed to humidity in matter of hours or days [18]. By using fluorescent doping in liquid resonators, however, we enable direct observation of WGMs without coupling to a tapered optical fiber and using a simple multimode laser for excitation of the dye, which simplifies the measurement process.

With respect to the material used for the fabrication of the liquid WGM RH sensor, it should be transparent in the excitation and emission wavelengths of the fluorescent dopant and show volumetric and refractive index changes with the RH along a wide range of RH, as well as stable properties in time and temperature, and in presence of chemical species existing in air.[6] In this work, glycerol (also known as glycerin) has been chosen because this polyalcohol is hygroscopic, that is, it attracts and absorbs water from the surrounding environment forming a stable solution. Moreover, this material is non-toxic, water-soluble, stable, viscous, and compatible with many other compounds.[19] Glycerol was mixed with a dispersion of rhodamine 6G in methanol in order to observe the shifts in the wavelength of the WGMs overlapped in the emission spectrum of this dye, due to the well-known Purcell effect, [20] which produces an increase in the emission at the resonant wavelengths. Using a similar technique, spectral tuning of glycerol-water-rhodamine B resonators has been already demonstrated, by electrical excitation.[21]

In WGM resonators, the resonant wavelengths follow the following relation:

$$m\lambda = 2\pi R \, n_{eff} \qquad (1)$$

Where an integer number *m* of resonant wavelengths $\lambda$, fits the circular optical path length in the resonator, which depends on its effective refractive index $n_{eff}$ and size $2\pi R$, being R the radius. Therefore, the shift in the wavelengths of the WGMs is a consequence of the change in the refractive index of the resonator as the glycerol absorbs/desorbs water from/to ambient depending on the RH. This effect effectively increases/decreases the radius of the droplet (the more water absorbed, the larger is the radius) and consequently decreasing/increasing the refractive index (because the refractive index of water is lower than that of glycerol)[19]. This behavior is depicted in **Figure 2b**.

With this work we aim to demonstrate the capability of glycerol doped with rhodamine 6G liquid WGM optical resonators as RH sensors using, for the first time, liquid resonators as ambient conditions sensors.

The liquid WGMs resonator characterized in this work (see **Figure 1b**) had an equatorial diameter of approximately 160 μm. The excited WGMs are located in a diameter perpendicular to the optical fiber that holds it.

In order to calibrate the droplet resonator as RH sensor, the emission of the rhodamine 6G had to be recorded (see the liquid resonator emitting in **Figure 1c**). An emission spectrum of the droplet resonator under excitation at 532 nm is plotted in **Figure 2a**. As expected, the emission band of rhodamine 6G (centered approximately at 600 nm) is modulated by the sharp peaks of the WGMs in the resonator.

The wavelength of the most intense peak of the rhodamine 6G emission (about 610 nm) was followed as the ambient RH was being slowly increased. As a result, the calibration shown in **Figure 3** was obtained.

A common figure of merit is sensitivity, *S*, defined for this particular sensor according to Equation 2 as the displacement of the wavelengths of the WGMs per RH percentage, relative to the wavelength value. Therefore, this value can be obtained as,

$$S = \frac{1}{\lambda}\frac{\delta\lambda}{\delta RH} \qquad (2)$$

As plotted in **Figure 4a** with a red line, the sensitivity of the calibrated droplet WGM sensor increases with RH, with a mean value about *2.8·10$^{-3}$ % RH $^{-1}$*. Other WGM resonators acting as RH sensors have achieved sensitivities of *2·10$^{-4}$ % RH $^{-1}$*, as did Zhang et al. using tubular hybrid optical microcavities.[22] Consequently, it can be said that, to the best of our knowledge, the droplet WGM sensor is one order of magnitude more sensitive to RH change than prior works with solid WGM RH sensors.

The relative displacement of the resonances can be estimated from experimental thermodynamical data[19] showing the equilibrium of the water + glycerol concentration for a given ambient RH and the refractive index of glycerol + water solutions for a given concentration and assuming that a small volume of the drop is occupied by the silica stem. The hygroscopicity of glycerol produces an increase in the radius and a decrease in its refractive index with the rise of the RH. As a result of this, the sensitivity of the displacement of the wavelengths of the WGMs with the RH can be obtained from Equations 1 and 2 as:

$$S = \frac{1}{\lambda}\left[\frac{1}{R}\frac{\delta R}{\delta RH} + \frac{1}{n_{eff}}\frac{\delta n_{eff}}{\delta RH}\right] \qquad (3)$$

Where R is the radius and $n_{eff}$ the effective refractive index of the droplet resonator. The sensitivity was found to be in good agreement with the result obtained from the experimental calibration, with a mean value of *S=2.5·10$^{-3}$ % RH $^{-1}$*, as shown in Figure 4a with a blue continuous line. The sensitivities of the two contributions (refractive index and radius) obtained from Equation 3 and thermodinamical data are also plotted individually in Figure 4a, with a blue dotted line for the sensitivity of the increase of the radius of the droplet with the RH and a blue dashed line for the sensitivity of the decrease of the refractive index of the liquid resonator with RH. Using this, the displacement of the wavelength of the WGM peak has been estimated in order to compare it with the experimental data of the calibration in Figure 3.

Due to the physics of this kind of sensor, there is no saturation in the extremes of the range of RH, as can be seen in the sensitivity (Figure 4a). At low RH, the shift is small but accordingly to Equation 1, since the denominator goes down, the sensitivity improves. At high RH, the shift is large, since the radius changes very fast due to the low concentration of glycerol in water.

Another parameter that could characterize the performance of the sensor is the resolution or limit of detection, that is, the magnitude of the minimum change in ambient RH that could be transduced, which can be calculated by means of Equation 3,

$$\Delta RH_{min} = \frac{\Delta \lambda_{min}}{\frac{d\lambda}{dRH}} \quad (4)$$

The common spectral shift sensitivity is about 1 % of the full width at half maximum of the WGM peaks.[17, 23] According to this, Ma et al. estimated for a coated silica microsphere acting as WGM RH sensor a detection limit of *$3 \cdot 10^{-3}$ % RH*.[17] However, quality factors of at least Q=$\lambda/\Delta\lambda_{min}$ = *$5 \cdot 10^5$* have already been experimentally demonstrated in liquid optical WGM resonators.[10] Having these facts into account in Equation (3), the droplet WGM RH sensor is expected to resolve a shift of *$7 \cdot 10^{-6}$ % RH*, besides the practical utility of such precision.

Regarding the dynamic of the process, the stabilization time can be estimated by using a random walk diffusion model given the diffusion coefficients D at room temperature and the penetration depth (which is reached in this time by the 68% of the molecules).

$$x_p = \sqrt{2 D t} \qquad (5)$$

The coefficient D at room temperature for water in silica is about $10^{-21}$ cm$^2$ s$^{-1}$ [24] and for water in glycerol is about $10^{-5}$ cm$^2$ s$^{-1}$. [25] By using the radius of the drop as the penetration depth we obtain a stabilization time of 3 s for the glycerol compared with $10^9$ years in silica. While this is relevant in order to achieve full sensitivity and stability, by reducing the size of the glycerol layer and increasing the diameter silica holder (neglecting diffusion in silica) the stabilization time can be shortened. In order to estimate the thinnest glycerol layer that can support WGM in the glycerol, a COMSOL simulation shown in Figure 4b displays the spatial distribution of the optical field (first radial mode, azimuthal 1200) in a perfect sphere. As can be seen, all the mode is within 2 μm from the surface. Using this as penetration depth, the stabilization time becomes 2 ms. In this scenario, the sensitivity decreases to the one associated to the refractive index change, that as can be seen in Figure 4a is about $10^{-2}$ per RH percent.

**Conclusions**

The optical WGM droplet resonator made out of glycerol, a hygroscopic liquid, shows a dependence of the refractive index and the radius of the resonator on the ambient relative humidity. Furthermore, rhodamine 6G is added to the resonator to enable non coupled observation of the WGM. It has a sensitivity of the displacement of the resonant wavelengths with environmental humidity of *2.8·10$^{-3}$* per relative humidity percent, and could achieve a resolution of *7·10$^{-6}$ %* of relative humidity.

**Experimental**

The liquid resonator consists of a droplet of glycerol and rhodamine 6G (with a 90 % by weight of glycerol) formed in the tip of a modified optical fiber, which is made by heating the end of a tapered optical fiber until it is slightly melted, so that superficial tension acts modifying its shape (see **Figure 1a**). This facilitates the formation of the droplet once the modified optical fiber is dipped into the liquid and also enhances its stability.[26] The presence of the optical fiber used to generate and manipulate the droplet originated slight deformations in the spherical form expected for the resonator (see Figure 1b) that had no significant influence on the resonances.[10]

The sensor was calibrated for the usual ambient RH range (approx. from 40 % to 65 % RH). The process was executed without any special insulation of the resonator and at room temperature. Before the calibration started, the ambient RH was lowered slightly below 40 % RH. Then, during the process, a controlled and steady increase of ambient moisture was held until 65 % RH was reached. Meanwhile, the ambient RH was measured using an electronic RH meter with a resolution of 0.5 % RH. At the same time, a set of measurements of the spectra of rhodamine 6G in the range from 550 to 650 nm were taken with a confocal microscope and pumping the rhodamine 6G at 532 nm with a commercial continuous wave solid state laser (see **Figure 1d**)[13].

**Acknowledgements**
This research was supported by MICINN (MAT2013-46649-C4-4-P), Consolider-Ingenio 2010 Program MALTA (CSD2007-0045), EU-FEDER, ICore: the Israeli Excellence center "Circle of Light", ISF (2013/15).

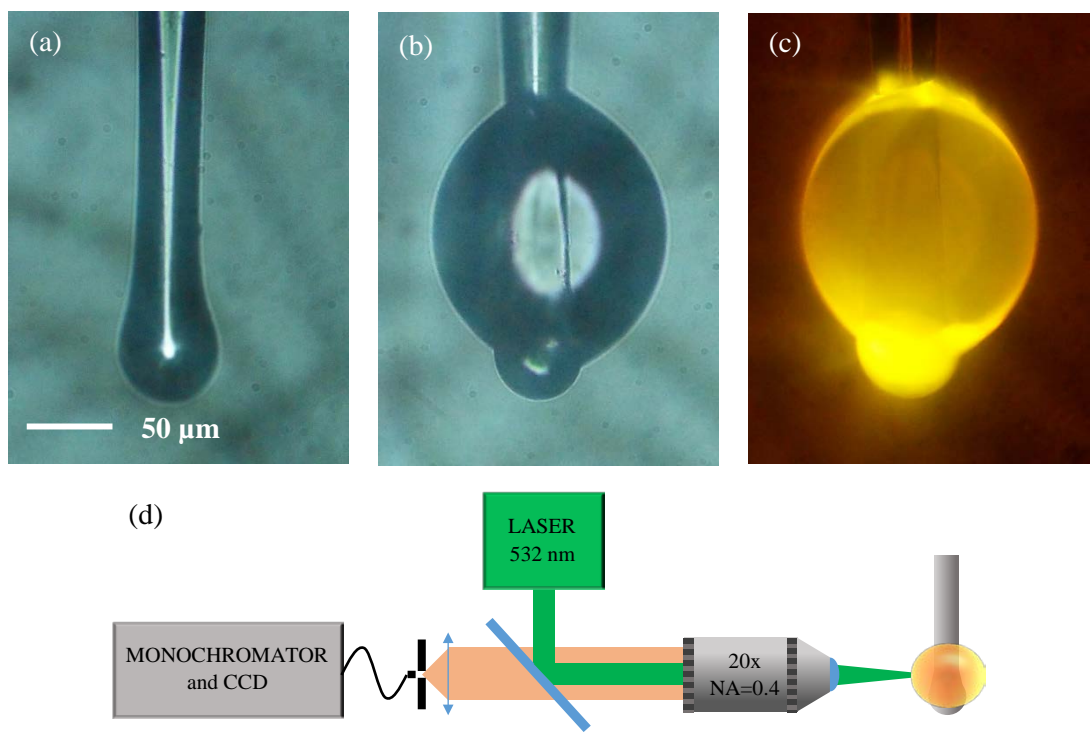

**Figure 1.** (a) Optical image of the modified optical fiber, (b) optical image of the liquid resonator made in the tip of the modified optical fiber, (c) optical image of the liquid resonator under excitation at 532 nm, and (d) schematic representation of the confocal microscope.

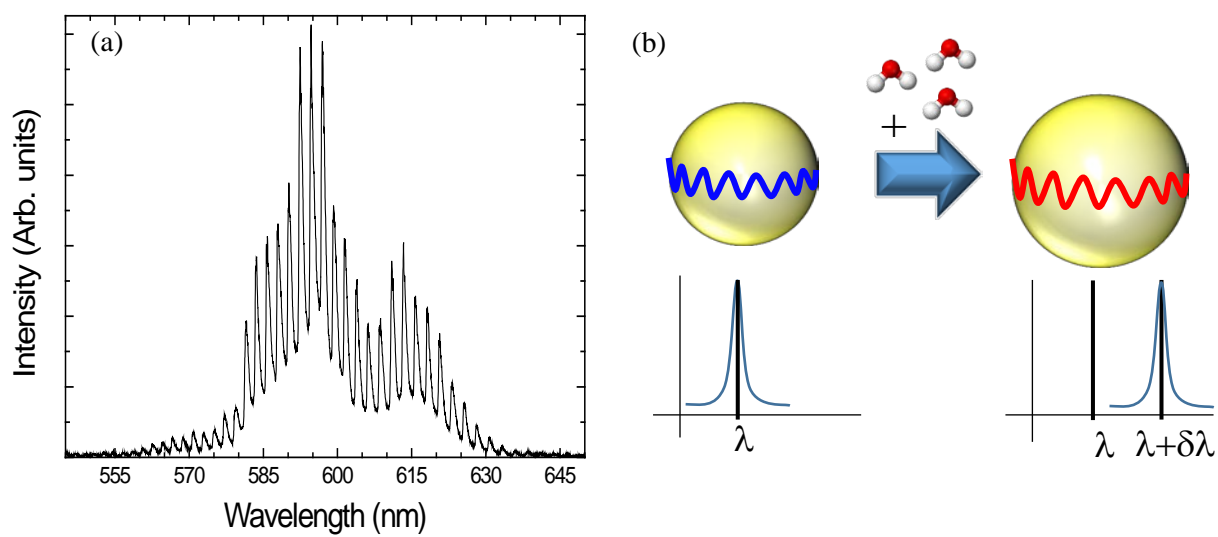

**Figure 2.** (a) Emission spectrum of rhodamine 6G present in the droplet resonator, where WGMs modulate the emission. (b) Scheme of the WGM shift due to water absorption in the drop.

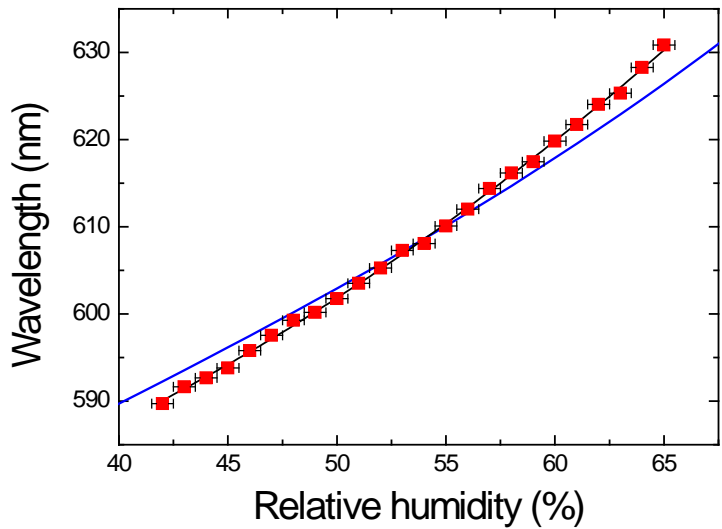

**Figure 3.** Calibration of the variation of the wavelength of a WGM peak with increasing RH (red squares) obtained from emission spectra of rhodamine 6G (see Figure 2), its fit to a quadratic polynomial (black line), and estimation of the displacement of the wavelength of the WGM peak from the expected variation of the refractive index and the radius of the resonator with RH (blue line).

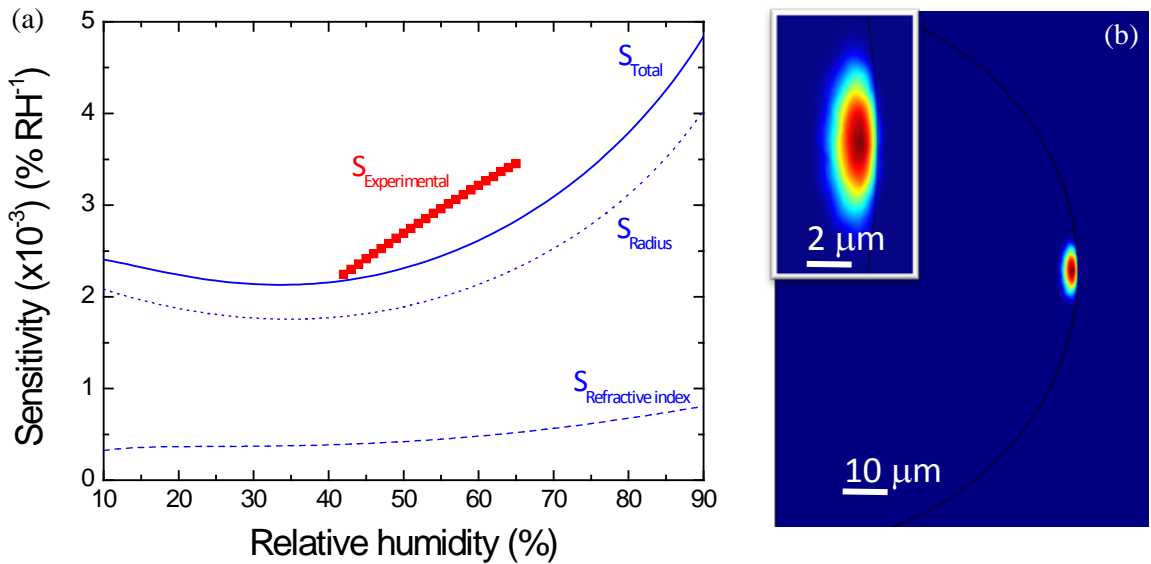

**Figure 4.** (a) Sensitivity of the displacement of the wavelengths of the WGMs calculated from the experimental data from the calibration using Equation (1) ($S_{Experimental}$, red squares) and from the estimation ($S_{Total}$, blue continuous line) of the sensitivity of the decrease of the refractive index ($S_{Refractive\ index}$, blue dashed line) and the sensitivity of the increase of the radius of the resonator ($S_{Radius}$, blue dotted line). (b) Numerical simulation of the first radial optical mode in a spherical droplet.